\documentclass[preprint,12pt]{elsarticle}

\usepackage{amssymb}
\usepackage{amsmath}

\usepackage[nolist]{acronym}
\usepackage{tabularx}
\usepackage{wrapfig}
\usepackage{multirow}
\usepackage{url}

\usepackage{subcaption}
\newcolumntype{M}[1]{>{\centering\arraybackslash}m{#1}}

\begin{acronym}[]
	\acro{AI}{Artificial Intelligence}
	\acro{ACC}{Adaptive Cruise Control}
	\acro{ADAS}{Advanced Driver Assistance System}
    \acro{API}{Application Programming Interface}
	\acro{CAN}{Controller Area Network}
	\acro{ECU}{Electronic Control Unit}
	\acro{GDPR}{General Data Protection Regulation}
	\acro{GOMS}{Goals, Operators, Methods, Selection rules}
	\acro{HCI}{Human-Computer Interaction}
	\acro{HMI}{Human-Machine Interaction}
	\acro{HU}{Head Unit}
	\acro{IVIS}{In-Vehicle Information System}
	\acro{KLM}{Keystroke-Level Model}
	\acro{LCA}{Lane Centering Assist}
    \acro{LDTM}{Lean Design Thinking Methodology}
	\acro{MGD}{Mean Glance Duration}
	\acro{OEM}{Original Equipment Manufacturer}
	\acro{OTA}{Over-The-Air}
	\acro{AOI}{Area of Interest}
	\acro{TGD}{Total Glance Duration}
    \acro{UI}{User Interface}
	\acro{UCD}{User-Centered Design}
	\acro{UX}{User Experience}
\end{acronym}

\journal{Transportation Research Part C}

\begin{document}

\begin{frontmatter}



\title{Predicting Multitasking in Manual and Automated Driving with Optimal Supervisory Control}




\author{Jussi P.P. Jokinen} 
\affiliation{organization={Cognitive Science, University of Jyväskylä}, 
            city={Jyväskylä},
            country={Finland}}

\author{Patrick Ebel} 
\affiliation{organization={ScaDS.AI, Leipzig University}, 
            city={Leipzig},
            country={Germany}}

\author{Tuomo Kujala} 
\affiliation{organization={Cognitive Science, University of Jyväskylä}, 
            city={Jyväskylä},
            country={Finland}}

\begin{abstract}
Modern driving involves interactive technologies that can divert attention, increasing the risk of accidents.  
This paper presents a computational cognitive model that simulates human multitasking while driving.  
Based on optimal supervisory control theory, the model predicts how multitasking adapts to variations in driving demands, interactive tasks, and automation levels.  
Unlike previous models, it accounts for context-dependent multitasking across different degrees of driving automation.  
The model predicts longer in-car glances on straight roads and shorter glances during curves.  
It also anticipates increased glance durations with driver aids such as lane-centering assistance and their interaction with environmental demands.  
Validated against two empirical datasets, the model offers insights into driver multitasking amid evolving in-car technologies and automation.

\end{abstract}




\end{frontmatter}



\section{Introduction}

Modern cars integrate various interactive technologies that provide information and entertainment to the driver but can also cause distractions.
Multitasking while driving is a major contributor to traffic accidents and near-accidents~\cite{Klauer.2014, Victor.2015}.
Emerging intelligent systems, particularly advanced driver assistance systems, aim to mitigate this risk by partially assuming vehicle control to reduce the driver's attentional demands.
How drivers adjust their attention and multitasking in response to different levels of automation directly affects the safety benefits these systems provide~\cite{bianchi2020drivers,dunn2021investigating,lin2018interview}.
With increasingly sophisticated automation and in-vehicle systems, understanding the complex relationship between driver multitasking, vehicle interaction design, and road safety is critical.

This paper presents a computational cognitive model that simulates how in-car multitasking adapts to the driving environment and the driver's internal cognitive processes.
The model captures multitasking under varying conditions, including changes in the driving environment, in-car interactive task design, and automation.
It is based on optimal supervisory control theory, which posits that humans allocate limited attentional resources across interleaved tasks according to a bounded-optimal solution~\cite{jokinen2021multitasking,jokinen2021touchscreen}.
Building on the concept of computational rationality~\cite{chandramouli2024workflow,gershman2015computational,lewis2014computational,oulasvirta2022computational}, we argue that behaviors such as multitasking while driving reflect optimal adaptations to constraints given specific objectives~\cite{jokinen2021multitasking, jokinen2021touchscreen}.
A model based on this perspective can predict behavioral changes as task conditions evolve, enabling model-based \textit{``what if''} scenarios and informing the design of interactive task environments.

\begin{figure*}[!t]
  \includegraphics[width=1\textwidth]{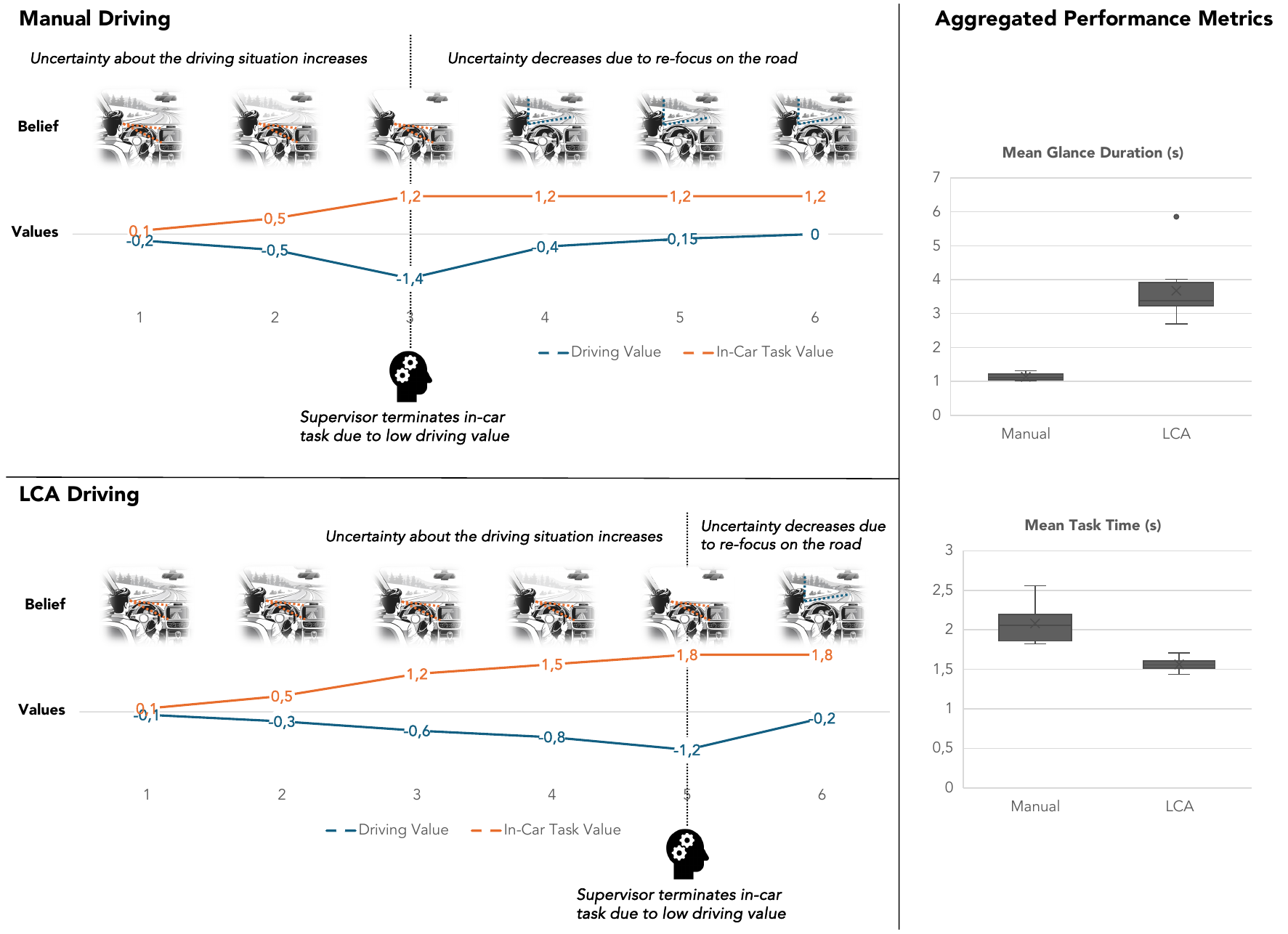}
  \caption{Our model of multitasking while driving simulates adaptation of task interleaving to internal cognitive constraints and the external environment.
    A multitasking ``supervisor'' allocates visual attention either to driving or an in-car visual search task based on expected joint task values.
    The top panel illustrates how an in-car glance results in increased uncertainty about the driving task, decreasing value predictions of the driving task.
    Finally, this compels the driver to terminate the visually demanding in-car task and return to the road, halting the progression of the in-car task.
    The bottom panel contrasts this with a situation where the lane centering assist (LCA) is turned on: due to the LCA helping to keep the lane, the value of the driving subtask decreases more slowly, resulting in a longer in-car glance.
    The boxplots on the right show aggregated results from our model: LCA allows for an adapted multitasking strategy that results in longer in-car glances and faster task completion.
  }
\label{fig:fig1}
\end{figure*}

Figure~\ref{fig:fig1} illustrates our model's capabilities by depicting sequential changes in a driver's internal cognition and multitasking behavior.
It compares two scenarios: one with the lane centering assist (LCA) activated and one with manual driving.
During in-car glances (e.g., glancing at the infotainment system), the driver relies on internal simulations to maintain a belief about the probable state of the driving environment.
However, noise from both the external environment and the driver's cognitive processes increases uncertainty in this estimate as the in-car glance duration increases.
In the top row of Figure~\ref{fig:fig1}, as the value estimates for the driving task decrease, those for an in-car search task (e.g., performed on the infotainment system) increase due to its progression.
Eventually, the driver terminates the in-car task and redirects visual attention to the road.
During LCA driving, uncertainty about the driving environment remains, but value estimates decrease more slowly because the driver has learned to tolerate greater uncertainty, modeling the driver's trust in automation.
As a result, predicted in-car glance durations are longer and task completion times shorter with LCA.

The model presented in this paper predicts outcomes as illustrated in Figure~\ref{fig:fig1} through a step-wise simulation, capturing transitions in driving task states and the driver's cognitive processes.
Our focus is on the driver's internal cognitive processes and their response to the external environment.
A detailed simulation of the driving environment is beyond the scope of this paper.
Nevertheless, we demonstrate that the model is not restricted to specific road settings, speed limits, or predefined levels of driving automation.
Instead, it generalizes an optimal multitasking rule set that adapts to various simulated road contexts and driving assistance systems.
The strategic allocation of visual attention between driving and in-car tasks emerges from internal cognitive constraints and their interaction with the external environment.

Our work complements previous simulation models of multitasking while driving and surpasses them in two key ways~\cite{gold2018modeling,jokinen2021modelling,jokinen2021multitasking,kujala2015modeling, Large.2017a, Purucker.2017, Ebel2023Forces}:
(1) it captures the adaptive nature of human cognition, and
(2) it expands the complexity of multitasking scenarios that can be simulated, deepening our understanding of the underlying cognitive processes.
Practitioners can use the model to examine how in-car interface design choices, here modeled as the complexity of the in-car search task, affect driver behavior and road safety.
Our contributions to computational cognitive modeling of multitasking while driving are as follows:

\begin{itemize}
\item We introduce a deep reinforcement learning model for multitasking in driving. It handles large, continuous state and action spaces, enabling realistic simulations of driving environments, including road curvature and dynamic adjustments to speed and steering.
\item Our model predicts driver multitasking behaviors under driving automation, such as lane-centering assistance.
\item We validate our model using two pre-existing empirical datasets:
\\
\begin{enumerate}
\item We evaluate the model's predictions of in-car glance durations based on driving speed and in-car interface design using a laboratory dataset. The model accurately predicts how driving speed, interface size, and task type affect glance durations.
\item We test the model’s predictions of in-car glance durations under driving automation using naturalistic data collected from human drivers interacting with in-car interfaces in real-world conditions. The model effectively captures how glance behavior adapts to lane assist at different speeds.
\end{enumerate}
\end{itemize}

\section{Background}

\subsection{Multitasking in Driving}

Driving inherently involves multitasking, requiring the simultaneous management of subtasks such as lateral and longitudinal vehicle control and hazard detection under a unified control process~\cite{Regan.2022,salvucci2009toward}.
In both manual and partially automated driving, drivers must respond to traffic and road changes, address sudden hazards, and monitor automation features~\cite{Regan.2022}.
While primary driving tasks impose a significant workload, drivers frequently engage in secondary tasks, such as interacting with infotainment systems or smartphones~\cite{ebel2023multitasking}.
These tasks compete for limited attentional resources and can impair the driver’s ability to respond to dynamic driving conditions.
Visually intensive tasks, especially those involving touchscreens, require drivers to look away from the road~\cite{Klauer.2014, Victor.2015, Ebel2023Forces} and pose a serious risk to road safety~\cite{Regan.2022}.

Although drivers adjust their multitasking in response to traffic conditions, driving situations, in-car interface design, and automation levels, their ability to adapt effectively depends on their certainty about the changing task environment~\cite{Young.2010, Rudin-Brown.BehaviouralAdaptation.2013, Oviedo-Trespalacios.2019, Oviedo-Trespalacios.2018a, Tivesten.2014, ebel2023multitasking, Morando.2019}.
To improve road safety and design less distracting \acp{IVIS}, a deeper understanding and modeling of drivers' multitasking behavior and adaptation to task environment changes are essential.
Below, we survey computational theories of driver multitasking on the basis of  Michon's driver task model, which categorizes decision-making into strategic, tactical, and operational levels, with increasing complexity from operational to strategic decisions ~\cite{Michon.CriticalView.1985, ebel2023multitasking}.

\subsection{Computational Models of Drivers' Multitasking}

A key objective of computational modeling of driver multitasking is to simulate human adaptation and predict multitasking behavior amid the uncertainty of distributing limited attentional resources across competing tasks.
Beyond improving our understanding of cognition and design interactions, computational models can potentially reduce the need for extensive and costly empirical testing with human drivers~\cite{murray2022simulation, Ebel.Narrowing.2021}.
In HCI and human factors research, modeling aims to predict the effects of in-car UI designs on driver behavior~\cite{lorenz2024computational,jokinen2021multitasking,Large.2017a, Purucker.2017, Ebel2023Forces}, evaluate the impact of driving automation and take-over requests~\cite{gold2018modeling,jokinen2021modelling}, and analyze interactions with other road users, such as pedestrians~\cite{wang2023modeling,zgonnikov2022should}.

\paragraph{Theory-driven models vs. data-driven models}

Models of human multitasking include \textit{theory-driven} cognitive models, such as multiple resource theory~\cite{wickens2002multiple}, SEEV~\cite{Sheridan1970, wickens2003SEEV}, threaded cognition~\cite{salvucci2008threaded}, and optimal control~\cite{jokinen2021multitasking,jokinen2021touchscreen}, as well as \textit{data-driven} models based on machine learning or probabilistic modeling~\cite{aksjonov2018detection, Ebel2023Forces, liu2015driver, mcdonald2020classification}.
Theory-driven models ground predictions in psychological hypotheses about human multitasking mechanisms, while data-driven models identify behavioral patterns directly from driver data.
Theory-driven approaches are valuable for understanding multitasking behavior because they explicitly incorporate cognitive mechanisms into model construction.

Although methods like LIME~\cite{Ribeiro.2016} and SHAP~\cite{Lundberg.2017} improve the explainability of black-box data-driven models, particularly in machine learning, they only provide predictions.
Unless a data-driven model is constrained to behave in a human-like manner, there is no guarantee that it accurately represents human behavior or that its explanations reflect actual cognitive processes.
Theoretical models are argued to generalize better by incorporating hypotheses about fundamental cognitive processes, whereas statistical models may struggle to extend beyond their training data, especially when the task environment changes significantly~\cite{murray2022simulation,oulasvirta2019s,oulasvirta2022computational}.
Conversely, data-driven models can uncover behavioral patterns from large datasets without requiring explicit hypotheses about the underlying cognitive mechanisms.
In this paper, we focus on cognitive modeling grounded in evidence-based psychological hypotheses about human multitasking.

In classical theory-based modeling approaches such as GOMS~\cite{Card.1983}, KLM~\cite{Card.1980}, and ACT-R~\cite{anderson2004integrated}, modelers define a set of instructions or \emph{productions} that specify how tasks are performed within the human cognitive and motor systems.
The model executes these instructions, simulating cognitive processes such as memory retrieval, visual attention, and motor actions to predict observable task behavior.
For example, ACT-R has been used to simulate how drivers visually interact with in-car interfaces~\cite{kujala2015modeling}.
While this improves our understanding of how interface design affects multitasking, production-rule-based models do not inherently predict how drivers adapt multitasking to varying driving conditions or designs~\cite{lorenz2024computational}.
This limitation arises because modelers may not anticipate behavioral adaptations, preventing their inclusion in the model's productions~\cite{oulasvirta2022computational}.
The need to manually define production rules or operator sequences further constrains the broad applicability of these classical theory-driven models.
Yet, predicting such adaptations is essential for understanding multitasking in complex, dynamically changing environments~\cite{jokinen2021multitasking}.

\paragraph{Computationally Rational Models}

Models based on computational rationality are gaining recognition in cognitive science and HCI~\cite{oulasvirta2022computational}.
The core principle is that human cognition adapts optimally to task constraints and its own limitations~\cite{gershman2015computational,lewis2014computational,oulasvirta2022computational}.
Although humans do not behave perfectly rationally by external standards, computational rationality models behavior as boundedly optimal within given constraints.
A recent model suggests that multitasking while driving is an optimal response to uncertainties arising from divided visual attention between primary driving tasks and secondary in-car activities~\cite{jokinen2021multitasking}.
During in-car glances, drivers maintain a \textit{belief} about the driving task, but uncertainty in this representation increases with glance duration.
At a critical point, considering the competing objectives of safe driving and completing the in-car task -- and factors such as driving speed -- redirecting gaze to the road becomes the optimal choice.
This theory does not imply that multitasking while driving is inherently rational, best, or safest.
Rather, it suggests that when drivers engage in in-vehicle tasks, even with potential crash risks, they adapt their multitasking optimally within the constraints of the driving environment, cognitive capacity, and motor functions.

The computational rational model of multitasking proposed by \citet{jokinen2021multitasking} employs hierarchical control, where a supervisory controller allocates visual attention among subtask controllers. This hierarchy consists of three \emph{agents} (controllers) that adjust their \emph{policies} based on specific goals: safe driving, efficient in-car search, and maximizing overall task performance (i.e., safe driving \emph{and} efficient search)~\cite{jokinen2021multitasking}.
Despite its simple structure, the model predicts how drivers adjust in-car glance durations in response to factors such as secondary task complexity, driving speed, and \ac{LCA}~\cite{jokinen2021modelling}.
However, it remains largely theoretical, assuming straight or slightly curved roads and constant speeds, without incorporating complex steering and speeding dynamics, which limits its applicability in more complex simulations.
While the model captures multitasking behavior using fundamental principles, its simplistic internal representation of driving environments constrains its generalizability.
In this paper, we address these limitations.

\subsection{The Goals for This Paper}  

Our review highlights driver adaptation as a key cognitive phenomenon that must be modeled to better understand how driver multitasking interacts with dynamic task environments.  
While recent efforts based on optimal supervisory control theory have provided initial insights, they do not fully capture cognitive adaptation in complex environments.  
These environments involve variations in speed and road configurations that interact dynamically with the surroundings and allow us to model the affect that driver assistance systems have on driver behavior. As automation and human-AI collaboration play an increasingly important role, the ability to model human adaptation to automation extends beyond driving alone. 

Table \ref{tab:model_comparison} provides an overview of the essential adaptive capabilities a computational cognitive model must account for to replicate human-like driver behavior, alongside a comparison of existing models.  
We consider only models that (1) aim to simulate human-like behavior, excluding autonomous driving models, and (2) perform the primary driving task, including both longitudinal and lateral control, alongside a multitasking component.  
This excludes models that predict multitasking behavior while driving but do not simulate the driving task itself (e.g., \cite{Ebel2023Forces, Purucker.2017, Large.2017a}).  

\begin{table}[]
  \small
  \setlength{\tabcolsep}{4pt}  
\centering
\caption{Comparison of different driving models in terms of what human-like adaptive phenomena they are able to model.}
\label{tab:model_comparison}
\begin{tabular}{m{0.15\linewidth} m{0.25\linewidth}M{0.08\linewidth} M{0.08\linewidth} M{0.08\linewidth} M{0.08\linewidth} M{0.06\linewidth} M{0.06\linewidth}}

\hline
\textbf{Level} & \textbf{Task} & This & Jokinen & Salvucci & Wortelen & Liu & Janssen \\
 &  & Model &  \cite{jokinen2021multitasking} & \cite{salvucci2006modeling} & \cite{Wortelen2013} & \cite{Liu2006} & \cite{janssen2010strategic}
   \\ \hline
\multirow{4}{*}{\begin{tabular}[c]{@{}l@{}} \textbf{Strategic} \end{tabular}}
& Goal setting  &  &  &  &  &  &  \\
& Route choice &  &  &  &  &  &  \\
& ADAS choice &  &  &  &  &  &   \\
& Multitasking choice & $\bullet$ & $\bullet$ & & $\bullet$  &  &  \\ \hline
\multirow{9}{*}{\begin{tabular}[c]{@{}l@{}} \textbf{Tactical} \end{tabular}}
& Attention allocation & $\bullet$ & $\bullet$ &  & $\bullet$ & $\bullet$ &  \\
& Turning & &  &  &  &  &  \\
& Overtaking &  &  &  &  &  &  \\
& Negotiation and signaling &  &  &  &  &  &  \\
& Hazard perception/prediction &  &  &  &  &  &  \\
& Speed acceptance & $\bullet$ &  &  & $\bullet$  &  &  \\
& Headway acceptance &  &  & $\bullet$ &  & &  \\
& Gap acceptance &  &  &  &  &  & \\
& ADAS usage & $\bullet$ &  &  &  &  &  \\ \hline
\multirow{3}{*}{\begin{tabular}[c]{@{}l@{}} \textbf{Operational} \end{tabular}}
& Critical response &  &  &  &  &  & \\
& Lateral control & $\bullet$ & $\bullet$ & $\bullet$ &$\bullet$ & $\bullet$ & $\bullet$\\
& Longitudinal control & $\bullet$ & & $\bullet$ & $\bullet$ &  &  \\ \hline
\end{tabular}
\end{table}

The categorization follows \citet{michon1985critical}.  
At the operational level, actions are controlled through automatic patterns and occur on a millisecond scale.  
At the tactical level, decisions unfold over seconds, while at the strategic level, decisions occur over longer intervals or can be made before driving.
Following items are listed in table, along whether existing models and ours consider them.

\begin{itemize}
    \item \emph{Goal setting (S)}: Prioritizing trip objectives, such as minimizing travel time, ensuring punctuality, maximizing safety, reducing effort, or optimizing fuel efficiency.  
    \item \emph{Route choice (S)}: Selecting roads, lanes, and intersections to reach the destination.  
    \item \emph{ADAS choice (S)}: Deciding whether to use driving assistance systems.  
    \item \emph{Multitasking choice (S)}: Choosing whether to engage in secondary tasks while driving.  
    \item \emph{Attention allocation (T)}: Distributing attention across different targets (e.g., roadway, other road users, \ac{IVIS}, speedometer, mirrors).  
    \item \emph{Overtaking (T)}: Changing lanes and accelerating to pass a lead vehicle.  
    \item \emph{Negotiating and signaling (T)}: Communicating with other road users (e.g., drivers, cyclists, pedestrians) to ensure efficient and safe traffic flow.  
    \item \emph{Hazard perception and prediction (T)}: Detecting and anticipating potential hazards requiring a timely response.  
    \item \emph{Speed acceptance (T)}: Selecting and adjusting speed based on road infrastructure requirements.  
    \item \emph{Headway acceptance (T)}: Maintaining a preferred time or distance headway to a lead vehicle.  
    \item \emph{Gap acceptance (T)}: Selecting a suitable gap to merge with intersecting traffic (e.g., at intersections, junctions, lane changes, pedestrian crossings).  
    \item \emph{ADAS usage (T)}: Selecting, adopting, and adapting available driver assistance systems. 
    \item \emph{Critical response (O)}: Executing automated reactions to sudden critical events to avoid collisions.  
    \item \emph{Lateral control (O)}: Steering to maintain lane position and avoid obstacles.  
    \item \emph{Longitudinal control (O)}: Adjusting speed via throttle and brake to maintain speed limits and preferred headway.  
\end{itemize}

This work aims to develop a unified theory of multitasking while driving, along with a model of the underlying cognitive processes. Ultimately, the goal is to create a predictive model that replicates human-like behavior in complex, real-world driving environments.
We contribute to the long-term goal of simulation modeling in HCI to provide practical tools that allow designers to evaluate, \emph{in silico}, how various design choices would theoretically affect driver multitasking~\cite{lorenz2024computational}.
To address these two goals, we 1)~present a formal theory of multitasking while driving, 2)~evaluate our model with two datasets of human driver multitasking, and 3)~discuss how our model could be designed for further realism to make it a tool to help design in-car interactions and automation.

\section{The Model}

\subsection{Optimal Control in Multitasking}

We model multitasking while driving as a hierarchical control problem, where the driver aims to maintain safe driving while allocating visual attention to an in-car task.
These concurrent tasks compete for shared cognitive resources, such as visual attention, creating a supervisory challenge: how to allocate these resources to benefit one subtask without compromising others too much.
The theory of optimal supervisory control suggests that this allocation is computationally rational: the supervisor evaluates alternative resource allocation policies and selects the one that maximizes the expected long-term joint subtask reward~\cite{jokinen2021multitasking,jokinen2021touchscreen,lingler2024supporting,oulasvirta2022computational}.

Our model architecture follows the structural analysis of computational rationality, distinguishing between an internal environment, an external environment, and an agent.
The \emph{internal environment} represents cognitive processes, while the \emph{external environment} describes the task environment~\cite{oulasvirta2022computational} (i.e., the driving environment and in-car task).
A representation of the external environment is constructed through observation and internal cognitive processing.
Based on this representation, the agent makes decisions, which in turn influence both cognitive processing and physical interactions with the environment.

We model the internal and external environments jointly as a Markov decision process (MDP).
An MDP is a tuple $<S, A, T, R>$, where:
\begin{itemize}
    \item $S$ is the (continuous) state space, representing all possible environmental states.
    \item $A$ is the (continuous) action space, encompassing all possible actions the agent can take.
    \item $T$ defines continuous state transition probabilities, given by $T: S \times A \times S \rightarrow [0,1]$, specifying a probability density function for transitioning to state $s' \in S$ after taking action $a \in A$ in state $s \in S$.
    \item $R$ is the reward function, defined as $R: S \times A \times S \rightarrow \mathbb{R}$, assigning a real-valued reward for transitioning from state $s$ to $s'$ via action $a$.
\end{itemize}
Additionally, we define a policy $\pi$ as a conditional probability distribution over actions given states: $\pi: S \rightarrow P(A)$, where $P(A)$ is the set of all probability distributions over the action space.

For a given policy $\pi$, the value function $V^\pi(s)$ represents the expected cumulative, time-discounted reward when starting from state $s$ and following policy $\pi$:
\begin{equation*}
  V^\pi(s) = \int_{A} \pi(a|s) \left( R(s, a, s') + \gamma \int_{S} P(s' | s, a) V^\pi(s') \, ds' \right) \, da,
\end{equation*}
where $\gamma \in [0,1]$ is a discount factor that weights future rewards.
Reinforcement learning (RL) seeks to derive an optimal policy $\pi^*$ that selects actions maximizing long-term cumulative discounted reward:
\begin{equation}
   V^{\pi^*}(s) = \max_{a \in A} \int_{S} P(s' | s, a) \left( R(s, a, s') + \gamma V^*(s') \right) \, ds' .
  \label{eq:bellman}
\end{equation}
These equations highlight the recursive nature of value estimation: an agent following a policy selects actions based not only on immediate rewards but also on expected future rewards, discounted over time, assuming continued adherence to the same policy.

In our hierarchical architecture, the driving and in-car subtasks are modeled as MDPs.
For the driving task, an optimal policy is derived using RL in continuous state and action spaces (Eq.~\ref{eq:bellman}), and for the visual search task, we use discrete state and action spaces.
The supervisor, also a separate RL agent, is trained with the converged (optimal) subtask agents.
Once the agents converge to optimal driving, search, and attention-sharing policies, forward simulations of multitasking, driving, and in-car interactions can be generated.
The following subsection details the construction of these components.

\subsection{The Supervisory Environment}

We model the allocation of visual attention between subtasks as an internal cognitive process that monitors their progress.
The supervisor observes the value $V^*(s)$ (Eq.~\ref{eq:bellman}) of both subtask agents—driving and in-car visual search—predicted from their respective optimal policies.
It does not access the actual state spaces of the subtask agents but relies solely on the values they compute based on their learned policies.

Since the subtask agents have learned their value functions through interaction with the environment, they can reliably predict the discounted cumulative long-term reward for a given state.
This enables the supervisor to learn an attention-sharing policy that maximizes the joint reward from both tasks.
The key intuition is that while the predicted value of the driving task decreases when attention shifts to the in-car task, negative rewards (e.g., from lane deviations) only occur after a violation has taken place.
Thus, the supervisor monitors the driving task value and learns a policy that redirects visual attention back to the road to minimize such future penalties.
Additionally, it tracks which subtask is currently receiving visual attention.

The supervisor receives its reward as the sum of the subtask rewards at each time step.
One step in the driving environment is defined as $0.1\,s$.
When visual attention remains on the driving task, the driving environment advances by one step.
If the supervisor shifts attention to the in-car search task, the driving environment advances twice ($0.2\,s$) without visual attention, accounting for the duration of the eye movement~\cite{kujala2015modeling, Sparks2002}.
The in-car search task then progresses by one step, followed by the driving environment advancing for the duration of that search step, which depends on the details of the interaction.
If attention shifts back to the road, the driving environment again advances twice without visual attention before resuming normal progression.
Since a single supervisor step may involve multiple driving environment steps, the accumulated rewards from the driving task (along with any from the search task) are pooled and returned to the supervisor for feedback during training.

\subsection{The Driving Environment}

\subsubsection{The Kinematic Model}

We simulate the driving environment using a kinematic bicycle model~\cite{gillespie2021fundamentals}.
While this model simplifies real-world driving by merging the front and rear wheels into single points, it effectively captures the impact of steering on vehicle movement and permits simulating features of driving that are relevant for developing our multitasking model.
At each time step $t$, the front wheel is located at $(x^f_t, y^f_t)$ and the rear wheel at $(x^r_t, y^r_t)$ on a 2D flat driving surface.
The wheels are separated by a fixed wheelbase of 2 meters.
The car's speed and heading are defined as $v \in [0, v_{max}]$ and $\theta \in [-\pi, \pi]$, respectively.
The driver controls the steering angle $\delta \in [-\delta_{max},\delta_{max}]$ and modulates acceleration or deceleration via $\mu \in [-\mu_{max},\mu_{max}]$.

In our simulation, the maximum speed is set to 150 km/h,
the maximum steering angle to $\delta_{max} = |0.26|$ radians,
and the maximum acceleration to $\mu_{max} = |6|$ m/s$^2$~\cite{fitch2010driver}.
Steering is affected by two noise sources: action-dependent noise $\sigma_\delta$ and action-independent noise $\sigma_h$.
This results in a noisy steering angle $\delta_\sigma$, defined as:
\begin{equation*}
  \delta_\sigma = \delta + |\delta| r_1 + r_2,
\end{equation*}
where $r_1$ and $r_2$ are normally distributed random variables with zero mean and standard deviations $\sigma_\delta$ and $\sigma_h$, respectively.
To mitigate instability at high speeds, a dampening factor $D$ is introduced~\cite{gillespie2021fundamentals,sharp2000mathematical}.
This factor decreases with vehicle speed, reducing noise effects:
\begin{equation}
  D = \left(1 - \frac{v}{v_{max}}\right).
  \label{eq:damp}
\end{equation}
Accordingly, the noisy steering terms are adjusted as $r_2 = r_2 \cdot D$ and $r_1 = |\delta| r_1 \cdot D$.
Finally, $\delta_\sigma$ is constrained within the car’s physical steering limits, ensuring $\delta_\sigma \in [-\delta_{max}, \delta_{max}]$.

Next, the car updates the positions of its front and rear wheels:
\begin{align*}
  x^f_{t+\Delta t} &= x^f_t + v \Delta t \cos(\theta + \delta_\sigma), \\
  y^f_{t+\Delta t} &= y^f_t + v \Delta t \sin(\theta + \delta_\sigma), \\
  x^r_{t+\Delta t} &= x^r_t + v \Delta t \cos(\theta), \\
  y^r_{t+\Delta t} &= y^r_t + v \Delta t \sin(\theta),
\end{align*}
where $\Delta t$ is the simulation time step ($0.1$ s) and $\theta$ is the car's heading.
Before updating, speed $v$ is modified by the control variable $\mu$: $v = v + \mu$.

The car's new heading is determined by the angle between the updated front and rear wheel positions:
\begin{equation*}
\theta_{t+\Delta t} = \text{atan2}(y^f_{t+\Delta t} - y^r_{t+\Delta t}, x^f_{t+\Delta t} - x^r_{t+\Delta t}).
\end{equation*}

The car's center position $(x, y)$ is computed as the average of the front and rear wheel positions.
To determine transitions between on-road and off-road states, we use an augmented Bresenham algorithm.
If the computed distance falls below a set threshold, a more precise distance from the car's center to the off-road location is calculated.

The kinematic bicycle model, despite its widespread use in vehicle dynamics studies, has limitations in realism~\cite{gillespie2021fundamentals,snider2009automatic}.
First, by representing vehicles as two-wheeled systems, it neglects roll, pitch, and yaw dynamics.
Second, it does not account for tire dynamics, such as deformation and slip.
Third, it overlooks aerodynamic forces and moments, which become significant at high speeds.
Additionally, the model excludes the effects of the powertrain and suspension on vehicle dynamics.
It also assumes perfect weight distribution and level terrain, limiting its accuracy in real-world and off-road scenarios.
Nevertheless, we chose this model for its simplicity and computational efficiency. Since our study prioritizes multitasking simulation over highly realistic driving dynamics, it provides a computationally efficient yet sufficiently accurate representation of steering effects.
In the final section, we discuss more realistic alternatives for future improvements to our cognitive model.

\subsubsection{Driving Assistance}

The simulation includes a feature that models the lance centering assist (LCA) as found in modern cars and actively adjusts the steering angle $\delta$ to keep the car in the center of the lane. 
The heuristic is straightforward: if the car nears a lane boundary, it steers sharply in the opposite direction.
Otherwise, it calculates distances to the left and right lane boundaries ahead and applies a minor steering adjustment to equalize them, thereby maintaining lane centering.
This heuristic-based LCA is not flawless, particularly under high steering noise or when driver inputs counteract its corrections.

\subsection{The Driver's Cognitive Environment}

The driver's cognition is modeled as an internal environment that maintains a belief about the driving task state and generates a reward signal based on internal preferences.
This environment consists of observation, belief update, and reward functions.

\subsubsection{Observation}
\begin{figure*}[!t]
  \includegraphics[width=1\textwidth]{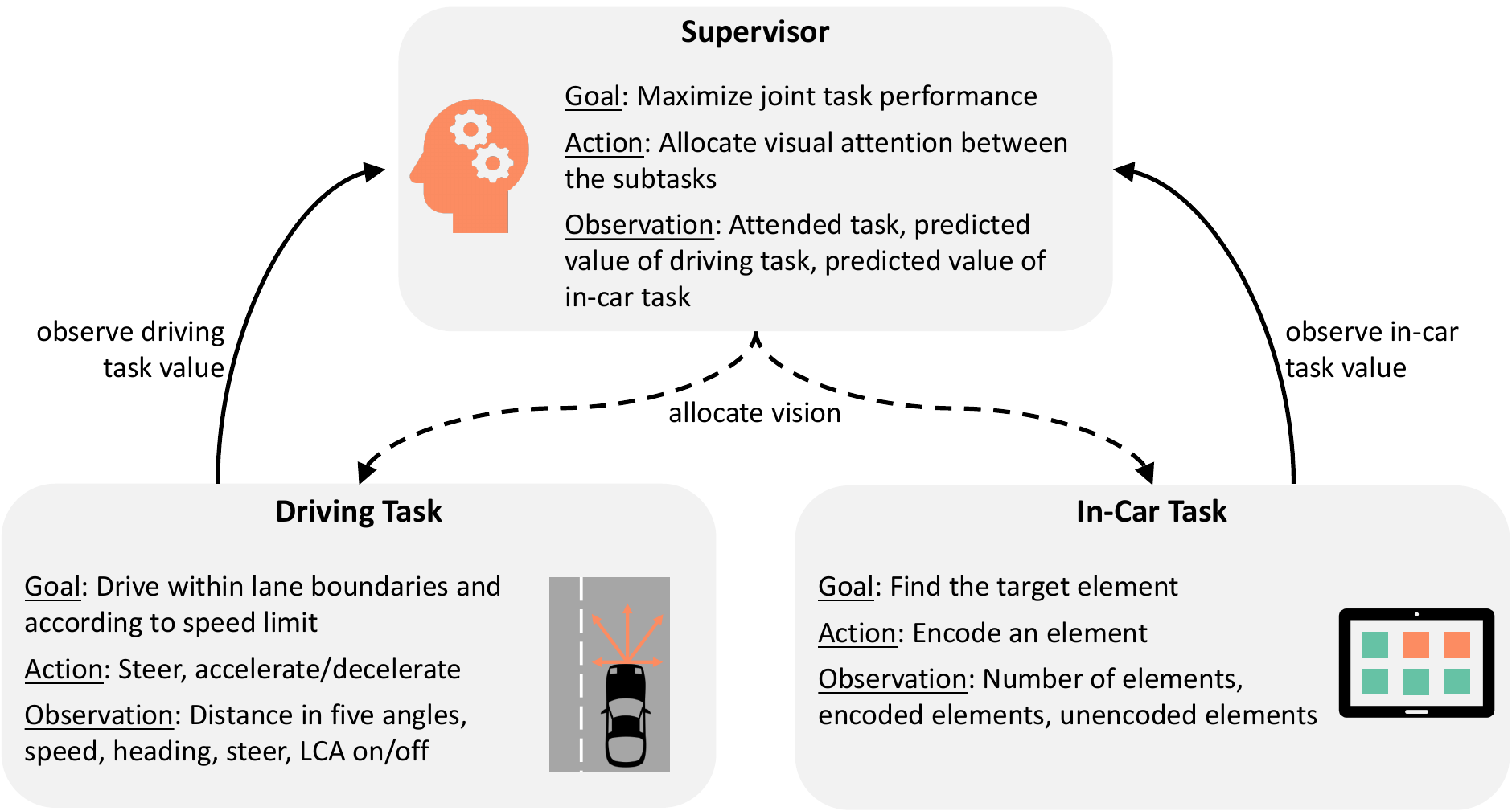}
  \caption{
    Our hierarchical multitasking model consists of a supervisor and two subtask agents.
    The supervisor allocates visual attention between the driving and in-car subtasks based on the value predictions of the subtask agents.
    These predictions reflect how well the current subtask states align with goal attainment. The supervisor aims to maximize long-term cumulative joint task performance.
  }
  \label{fig:belief}
\end{figure*}

When the simulated driver visually attends to the road, it makes a noisy observation of the driving environment.
The car's position $(x,y)$ is observed with independent Gaussian noise, where each coordinate is perturbed by a normal distribution with zero mean and standard deviation $\sigma_o$.
In contrast, speed, heading, and steering are observed without added noise.
In future iterations, all of the may include noisy observations, but more research would be needed to decide their relative weights.
Additionally, five distance metrics estimate proximity to lane boundaries based on the noisy position belief: directly ahead, 0.4 radians to the left and right, and 1.57 radians to both sides (see Figure~\ref{fig:belief}).
If the car is off-lane, these values represent the distances back to the lane in the specified directions and are assigned negative values to indicate an off-road status.

\subsubsection{Belief Update}

When the driver agent is not visually attending to the environment, the car's position $(x,y)$ is not observed.
Instead, it is estimated using an internal simulator, which is a cognitive physics engine that approximates environmental updates—based on the current belief.
This simulator predicts the next probable position $(x_b, y_b)$ but introduces noise in time estimates, sampled from a normal distribution $\mathcal{N}(\Delta_t, \sigma_t)$~\cite{taatgen2007integrated}.
Belief over the five lane boundary distances are computed based on this belief.
The driver has full knowledge of the static driving environment (its 2D map) but remains uncertain about the car's exact position within it, as the internal simulator does not account for noise affecting the car's steering, and introduces its own noise in position predictions due to noisy time step estimates.
As a result, the driver's belief state starts to diverge from the true state.

When the driver returns vision to the road, the positional belief $(x,y)$ is updated as a weighted sum of the internal prediction and the noisy observation.
The weights are computed based on their respective variances:
\begin{equation*}
w_p = \frac{\sigma_o^2}{\sigma_p^2 + \sigma_o^2}, \quad
w_o = \frac{\sigma_p^2}{\sigma_p^2 + \sigma_o^2},
\end{equation*}
where $\sigma_p$ is the dampened total noise.
The updated belief is then given by:
\begin{align*}
x &= w_p \cdot x + w_o \cdot x_b, \\
y &= w_p \cdot y + w_o \cdot y_b.
\end{align*}
This results in a filtered posterior belief that integrates both the internal predictive model and noisy observations.

To represent positional uncertainty, the cognitive environment maintains an estimate of uncertainty in the current position.
This uncertainty is updated using a simple linear model:
\begin{equation*}
\sigma^{pos}_{t + \Delta t} =
\begin{cases}
\sqrt{(\sigma^{pos}_t)^2 + \sigma_p^2 + \sigma_t^2}, & \text{if vision is absent}, \\
\sqrt{w_p^2 (\sigma^{pos}_t)^2 + w_o^2 \sigma_o^2}, & \text{otherwise}.
\end{cases}
\end{equation*}
This uncertainty representation is crucial for optimal supervisory control, as it allows the driver to predict declining values during prolonged visual inattention.

\subsubsection{Reward Function}

The agent receives a baseline reward of 0 when driving within the speed limits and staying on the road.
Violations incur penalties: a reward of -1 is applied for each time step the car is off-road.
If the vehicle exceeds or falls more than $10\,\text{km/h}$ beyond the speed limit, it receives a negative reward proportional to the speed difference, scaled by a penalty factor of -0.1.

The episode ends upon reaching the designated end area or is truncated after a threshold time if the agent fails to stay on the road during early training.
The objective is to develop a driving policy that minimizes penalties for off-road driving and speed violations, keeping the cumulative reward as close to zero as possible.
Model parameters are calibrated to ensure this performance is achievable with full visibility; however, during visual inattention, noise and imperfect belief updates may lead to driving errors even under an optimal policy.

During training, the agent is placed in a road environment and begins driving.
To learn driving without visual attention, the model's internal environment is randomly made visually inattentive for periods between 0.2 and 5 seconds.
As a result, the agent learns to associate increasing uncertainty in the driving environment with lower value expectations, which the supervisor will later learn to utilize in adapting an efficient multitasking policy.

\subsection{In-car Visual Search}

The in-car visual search task is modeled as a discrete search space, where each visual element represents an action to fixate on and encode.
Eye movement and encoding time are computed using the EMMA eye movement model~\cite{Salvucci2001}.
The task structure varies based on the number of rows and columns: it is completed either when a randomly located target is encoded (task type 0) or when all visual elements have been encoded (task type 1).

The search agent’s observation space includes a visual short-term memory buffer, which it learns to use to avoid revisiting previously fixated elements, as well as the current fixation location and interface size.
When the agent fixates on and encodes an element, it incurs a negative reward equal to the encoding time in seconds.
Successfully completing the search task yields a positive reward proportional to the number of elements.
The optimal policy of the visual search agent discovers inhibition of return by utilizing the visual short term memory structure, and minimizes search time by preferring short element-to-element saccades.

\subsection{Supervisor Progression}  

The multitasking model begins by resetting both the driving and search environments.  
The driving task starts at a speed equal to the speed limit, while the search agent is initialized with no observed elements and a randomly assigned target.  
The driving agent is first stepped with full visual attention for 10 steps (1 second) to establish a reliable belief, after which the supervisor takes control.  

At each time step, the supervisor selects a task for visual attention.  
If attention remains on driving, the driving environment advances one step, and its new value is recorded.  
If attention shifts to the in-car search task, the driving task proceeds unattended for $0.2\,s$, simulating the time required for eye movements to the in-car display~\cite{kujala2015modeling}.  
The search task then steps once, and its new value is recorded.  

While attention remains on the search task, the search environment advances one step per time step, while the driving task continues unattended for the duration of each search action (determined by the EMMA model).  
Once the search task is completed, the in-car display may continue to the next task, or, as in the case of our first experiment below, it may simulate task change by remaining inactive for a while before resetting.
If the supervisor shifts attention back to driving, the driving environment advances for the $0.2\,s$ without visual attention, followed by one step with full observability.

Due to noisy observations, the driver does not necessarily regain full situation awareness in a single step but requires multiple observation steps, which is reflected in the values recorded by the supervisor.  
Rewards from driving (potentially spanning multiple steps) and visual search are summed, with driving rewards weighted by $w_d=5$.  
This parameter is tuned to reflect the trade-off between maintaining safe driving and efficiently locating targets on the in-car display.  

The driving model employs the soft actor critic (SAC) algorithm~\cite{haarnoja2018soft}, with value predictions from the dual-Q policy network's minimum value prediction.
This choice was made given the continuous nature of the state and action spaces in the driving subtask's control problem.
The search agent is trained using Proximal Policy Optimization (PPO) \cite{schulman2017proximal}, with discrete observation and action spaces.
The supervisor, with its continuous observation space and discrete action space, also utilizes PPO.
Both algorithms are trained via \texttt{stable\_baselines3}, with all hyperparameters set to default, except for learning rate, which for all three agents is $0.0001$.

\section{Evaluation}

To evaluate our model, we first demonstrate its internal mechanics.
We then compare its predictions against data from a laboratory study on driver multitasking conducted in a driving simulator, while fine-tuning its free parameters.
Next, we assess the model's predictions using data from a naturalistic driving study.
Importantly, the model's parameters are not tuned for the naturalistic data; instead, it employs the parameters optimized with experimental data.
In fact, all simulations in this paper -- regardless of speed limit, road configuration, or driving automation -- use the same converged models for driving and supervision.

\subsection{Optimal Supervisory Control in Practice}

While the theory of optimal supervisory control has been established~\cite{jokinen2021multitasking,jokinen2021touchscreen}, there has yet to be a detailed exposition of how value observation and resource allocation between subtasks lead to human-like multitasking.
To illustrate this concept more concretely, we conduct a computational experiment to examine the core mechanisms of our model and the underlying theory.

First, we analyze the driver agent’s value predictions over time.
The agent drives with visual attention for approximately $2$ s, then without vision for a similar duration, before regaining visual attention.
For the optimal supervisory control model to function as intended, the agent should anticipate declining values during visual inattention, followed by a rebound upon re-focusing on the road.
Furthermore, at higher driving speeds, the agent should predict more negative values, reflecting the increased risk of visual inattention at faster speeds.
Finally, we assess the impact of LCA by comparing predicted values with and without automation.

\begin{figure*}[!t]
  \includegraphics[width=1.0\textwidth]{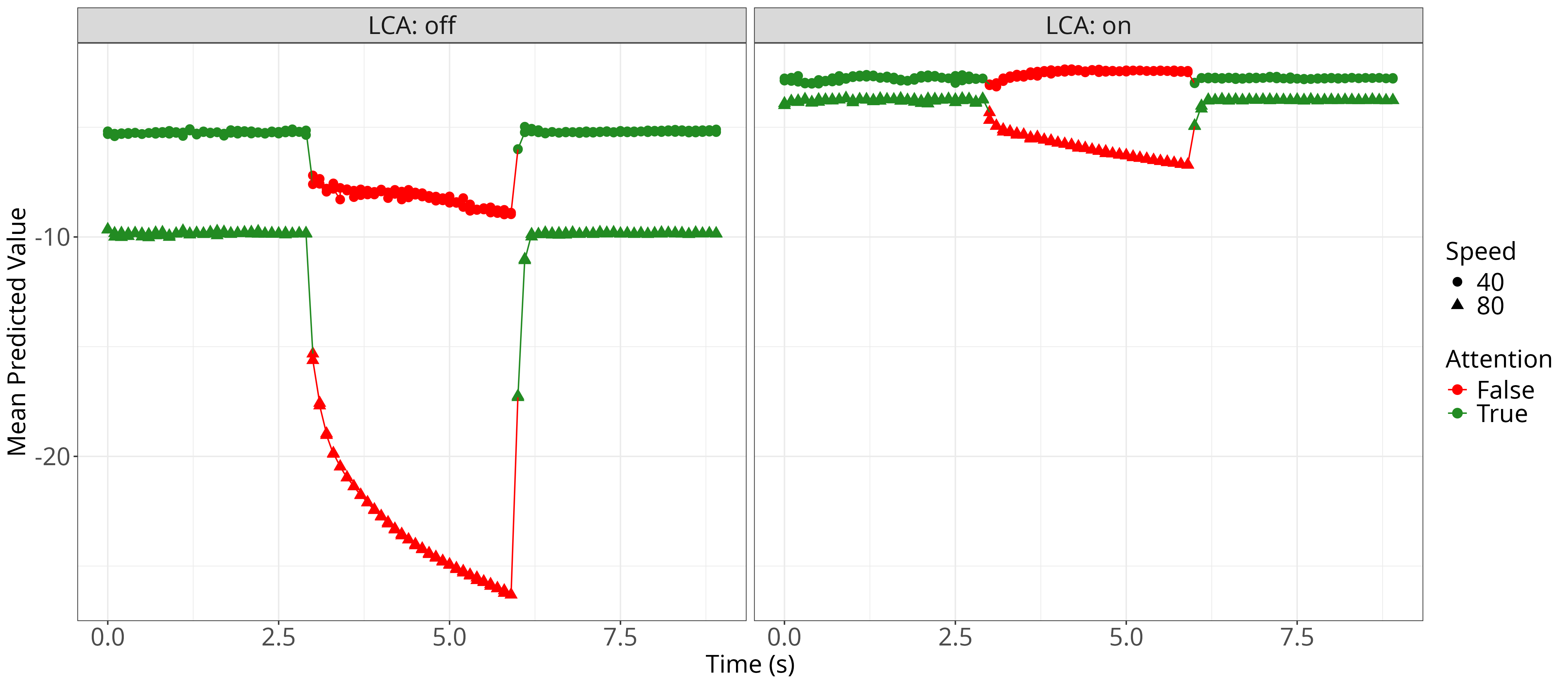}
  \caption{
    Changes in the driver agent's value predictions relative to visual attention and driving conditions.
    Predicted values decline during visual inattention (red) and recover upon re-engaging visual monitoring (green).
    The effects of driving speed (km/h) and car automation (LCA) are also visible.
    Notice the impact of LCA under the slower speed: the model predicts that its ability to drive solely under LCA might even be slightly better than with the driver with noisy vision intervening.
  }
  \label{fig:driver_value}
\end{figure*}

For clarity in presentation, we aggregate value predictions across multiple analogous episodes to account for inherent steering and observation variances within a single episode.
The results, shown in Figure~\ref{fig:driver_value}, support our hypothesis: visual inattention leads to a temporary decline in predicted values, followed by gradual recovery upon regaining visual attention.
Velocity influences these values, with an interaction effect: at higher speeds, the decline occurs more rapidly than at lower speeds.
Finally, when LCA is active, the agent predicts higher values, mitigating the negative impact of visual inattention.

\begin{figure*}[!t]
  \includegraphics[width=1.0\textwidth]{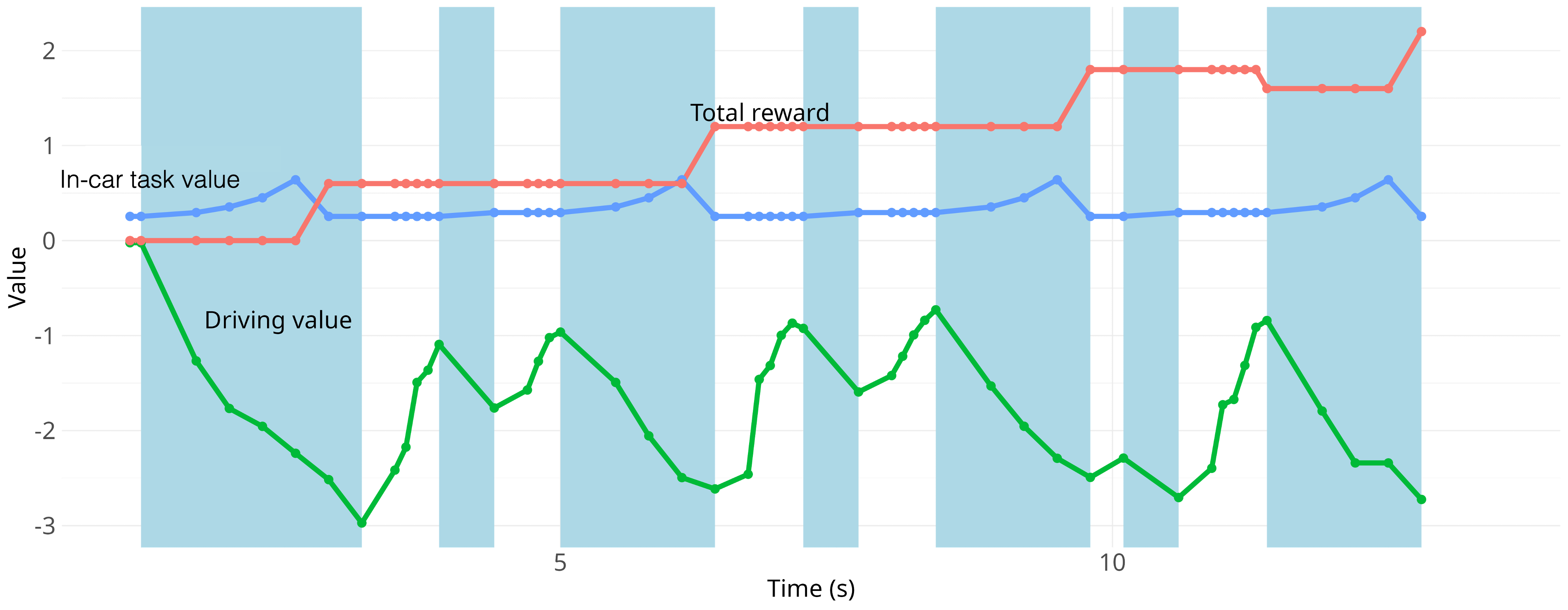}
  \caption{
    Predicted values of driving and in-car search subtasks (green and blue lines), cumulative reward collected in the episode (red line), and in-car glances (blue background shading) over time.
    The supervisor monitors dynamically changing subtask values and shifts visual attention accordingly.
    Driving values are vertically shifted upwards for improved visualization.
  }
  \label{fig:supervisor_value}
\end{figure*}

Next, we apply the full multitasking model to examine the core principle of optimal supervisory control: that supervisory decisions are based on observed subtask values.
Figure~\ref{fig:supervisor_value} illustrates the model’s multitasking trajectory over a 14 second driving period (values from the driver are vertically adjusted for clarity).
Initially, the supervisor shifts attention to the in-car search task (blue background), causing search values to increase with each search action while driving values decline sharply.
The supervisor maintains focus on the search task until it redirects visual attention back to driving (white background).
Once the supervisor agent allocates attention to the road, the value of the driving agent begins to recover, eventually leading the supervisor to shift attention back to the search task.

Figure~\ref{fig:supervisor_value} also shows the cumulative reward acquired by the supervisor, which increases upon search task completion but decreases around $t \approx 12$ s due to the car veering off-road.
Importantly, the supervisor does not observe these rewards during forward simulation (only during training) and thus cannot use them to determine when to shift visual attention.

\subsection{Predicting Multitasking while Driving}

The first dataset used to fit and validate the model comes from a medium-fidelity driving laboratory study reported in~\cite{jokinen2021multitasking}.
The dataset is openly available at https://gitlab.com/jokinenj/multitasking-driving.
In the study, human participants drove on a gently curved road at either 60 or 120 km/h while performing a visual search task on an in-car display.
Each search task required locating a target among distractors, with either 6 or 9 elements on the screen.
The target was either present or absent, and participants had to indicate whether they found it or determined it was not present.
Thus, the dataset includes average in-car glance durations and task completion times across 8 conditions (speed, number of elements, and task type).

To calibrate the model’s noise parameters for human-like driving, we adjusted them within a range that produced plausible behavior.
Excessive noise causes the driving agent to fail at lane-keeping even with full visual attention, while minimal noise makes prolonged visual inattention inconsequential.
Similarly, if internal time-keeping noise is too high, the agent quickly loses the ability to maintain a useful representation of the environment.

\begin{figure*}[!t]
  \centering
  \begin{subfigure}[t]{0.45\textwidth}
    \includegraphics[width=\textwidth]{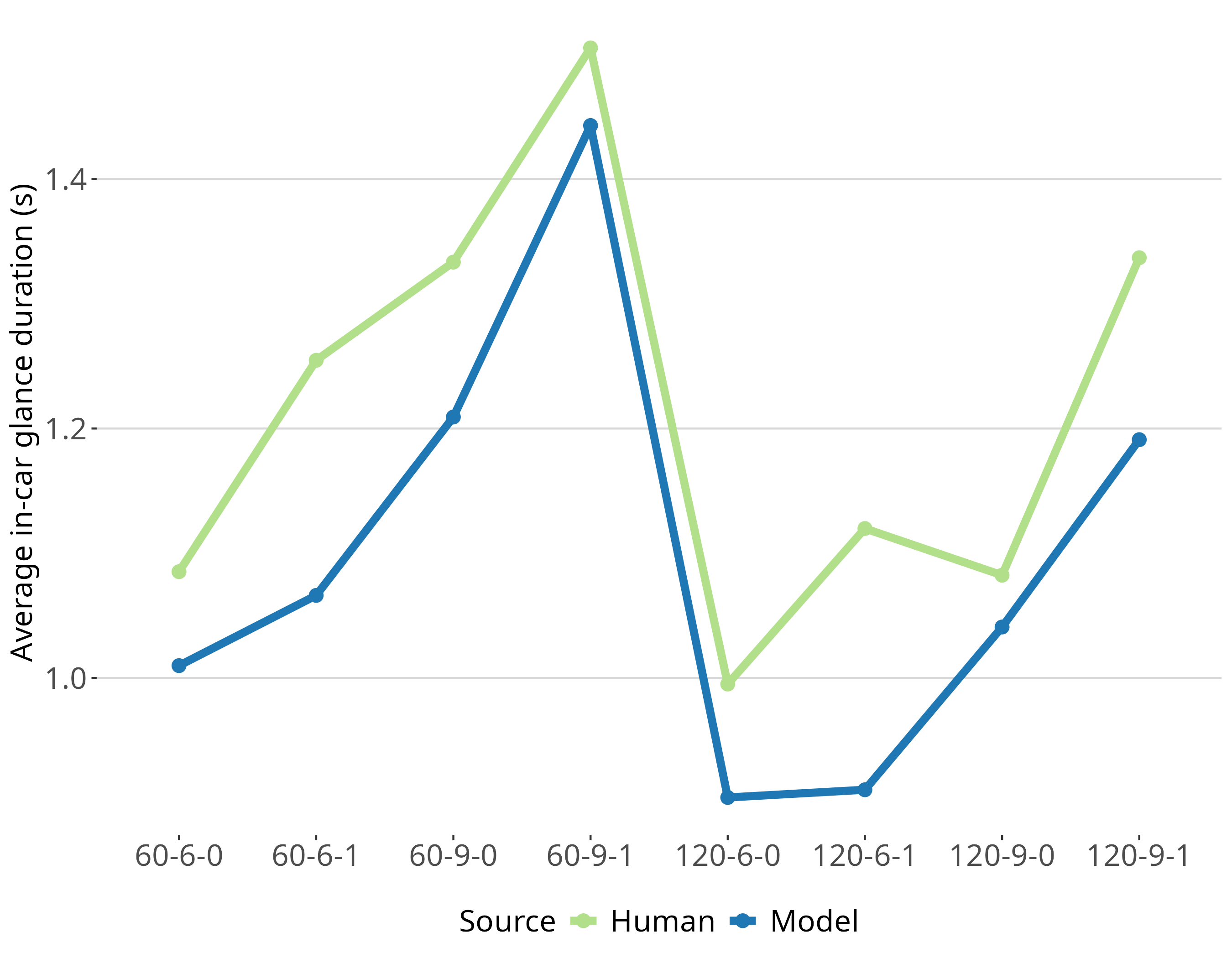}
    \captionsetup{justification=centering}
    \label{fig:fig_exp1a}
  \end{subfigure}%
  \begin{subfigure}[t]{0.45\textwidth}
    \includegraphics[width=\textwidth]{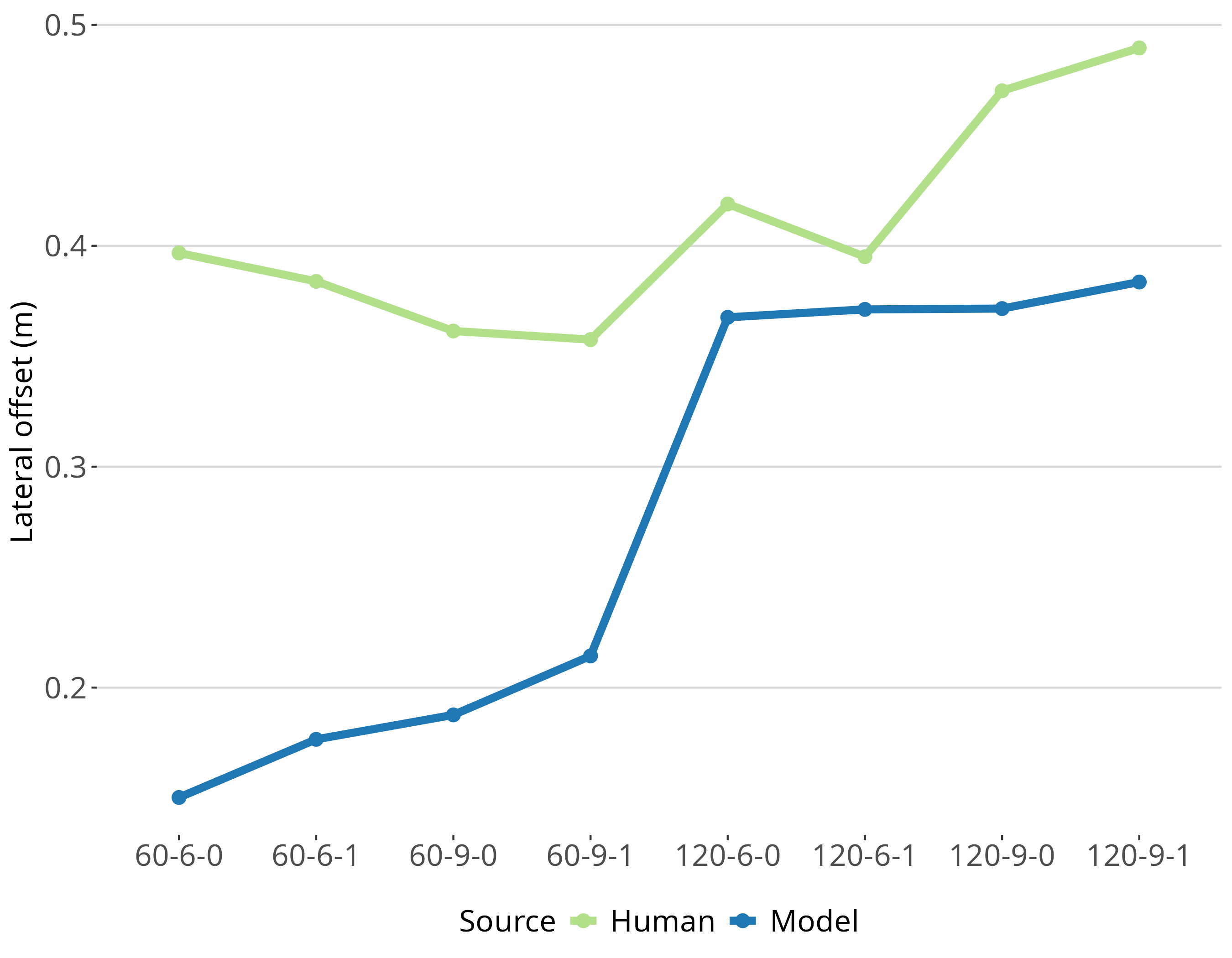}
    \captionsetup{justification=centering}
    \label{fig:fig_exp1b}
  \end{subfigure}%

  \begin{subfigure}[t]{0.45\textwidth}
    \includegraphics[width=\textwidth]{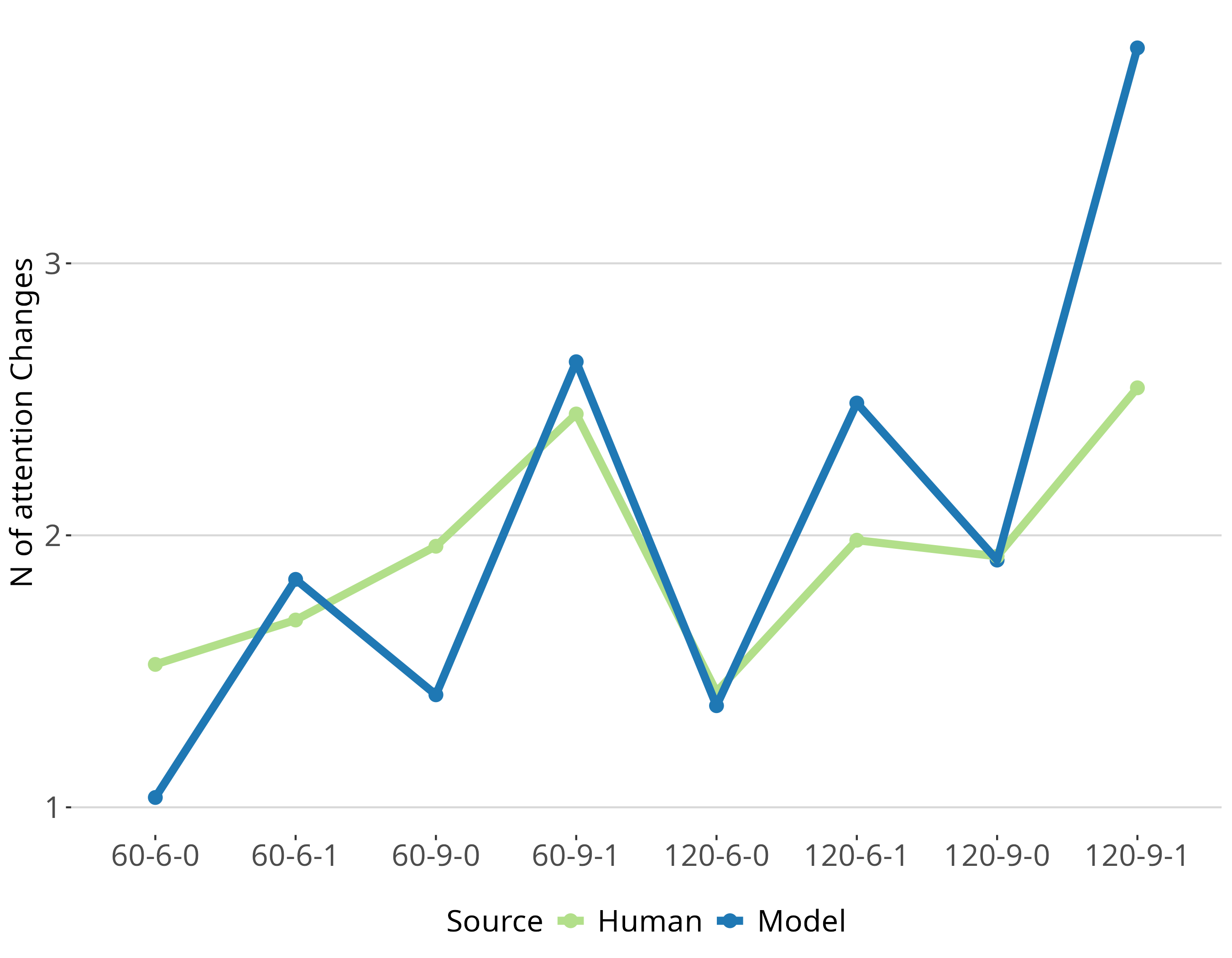}
    \captionsetup{justification=centering}
    \label{fig:fig_exp1c}
  \end{subfigure}%
  \begin{subfigure}[t]{0.45\textwidth}
    \includegraphics[width=\textwidth]{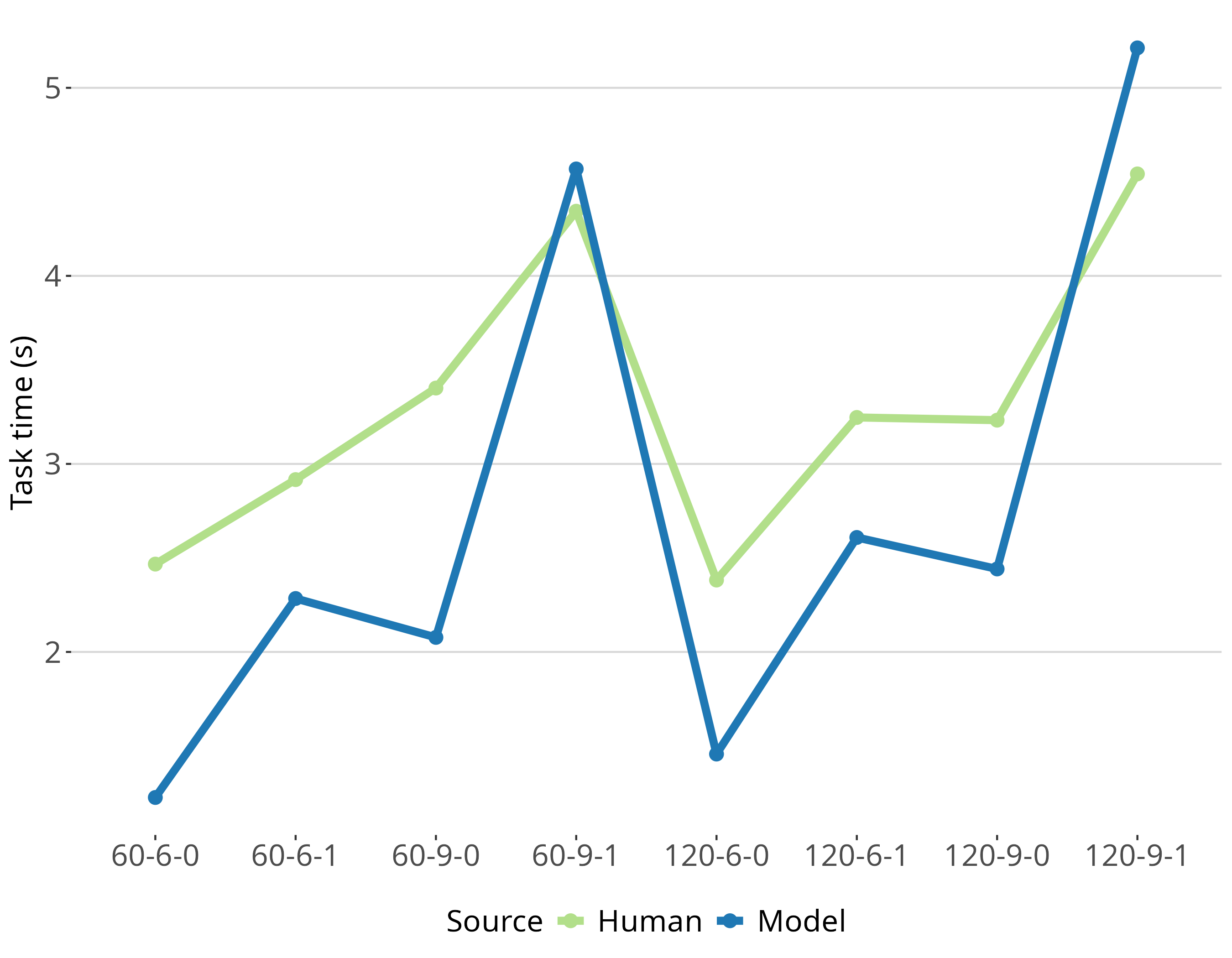}
    \captionsetup{justification=centering}
    \label{fig:fig_exp1d}
  \end{subfigure}%
  \caption{Comparison of our model's predictions and human data in key outcome variables. The x axes labels encode speed (60 or 120 km/h), number or items (6 or 9), and task type (0 = target on screen, 1 = no target on screen).}
  \label{fig:fig_exp1}
\end{figure*}

The final parameter values obtained are $\sigma_h = 0.0003$, $\sigma_\omega = 0.0001$, $\sigma_t = 0.0001$, and $\sigma_o = 0.0001$.
While we do not conduct rigorous flexibility testing on these parameters, the model does not appear highly sensitive to minor parameter variations.
Figure~\ref{fig:fig_exp1} compares the model’s predictions with human data.
The model successfully replicates the effects of experimental manipulations on key behavioral indicators, including average in-car glance duration, lateral offset, average in-car glances per task, and task duration.
Model fit for these indicators is as follows:
$R^2 = 0.89$, RMSE = 0.11;
$R^2 = 0.53$, RMSE = 0.39;
$R^2 = 0.77$, RMSE = 0.25;
$R^2 = 0.93$, RMSE = 0.30.
The model captures the influence of speed and in-car task design on these trends.
However, it occasionally underestimates lateral offset, possibly due to differences in road configuration.
For example, road width in the original human experiment varied slightly, and while human participants drove at a constant speed, our model allowed speed variations within an acceptable threshold.

\subsection{Predicting Naturalistic Driving}

To evaluate the model’s performance in more complex environments, we compare its predictions to data collected during naturalistic driving.
We use a dataset that has been partially reported in previous studies~\cite{ebel2023multitasking, Ebel2023Forces, Ebel2022HowDrivers}.
The data were collected over-the-air from more than 100 Mercedes-Benz test vehicles used for various test procedures, as well as employee commuting and recreational driving.
Spanning from February 2022 to February 2023, the dataset includes driving data (e.g., vehicle speed, steering wheel angle, automation status), eye-tracking data, and touchscreen interaction data.
No personal, demographic, or environmental information is included.
Signal-level processing followed the methodology described in~\citet{ebel2023multitasking}.

For evaluation, we compare model predictions only against manual driving sequences and sequences of LCA driving.
In these cases, adaptive cruise control also determined vehicle speed.
To align with the model’s speed limits, we exclude sequences where the average vehicle speed exceeded $150\,\text{km/h}$.
After processing and filtering, the final dataset for model evaluation consists of 27,676 interaction sequences from 7,511 individual trips.
The median sequence duration is 11.38\,s, and the median number of touchscreen interactions per sequence is 5.

\begin{figure}[!ht]
  \includegraphics[width=0.8\textwidth]{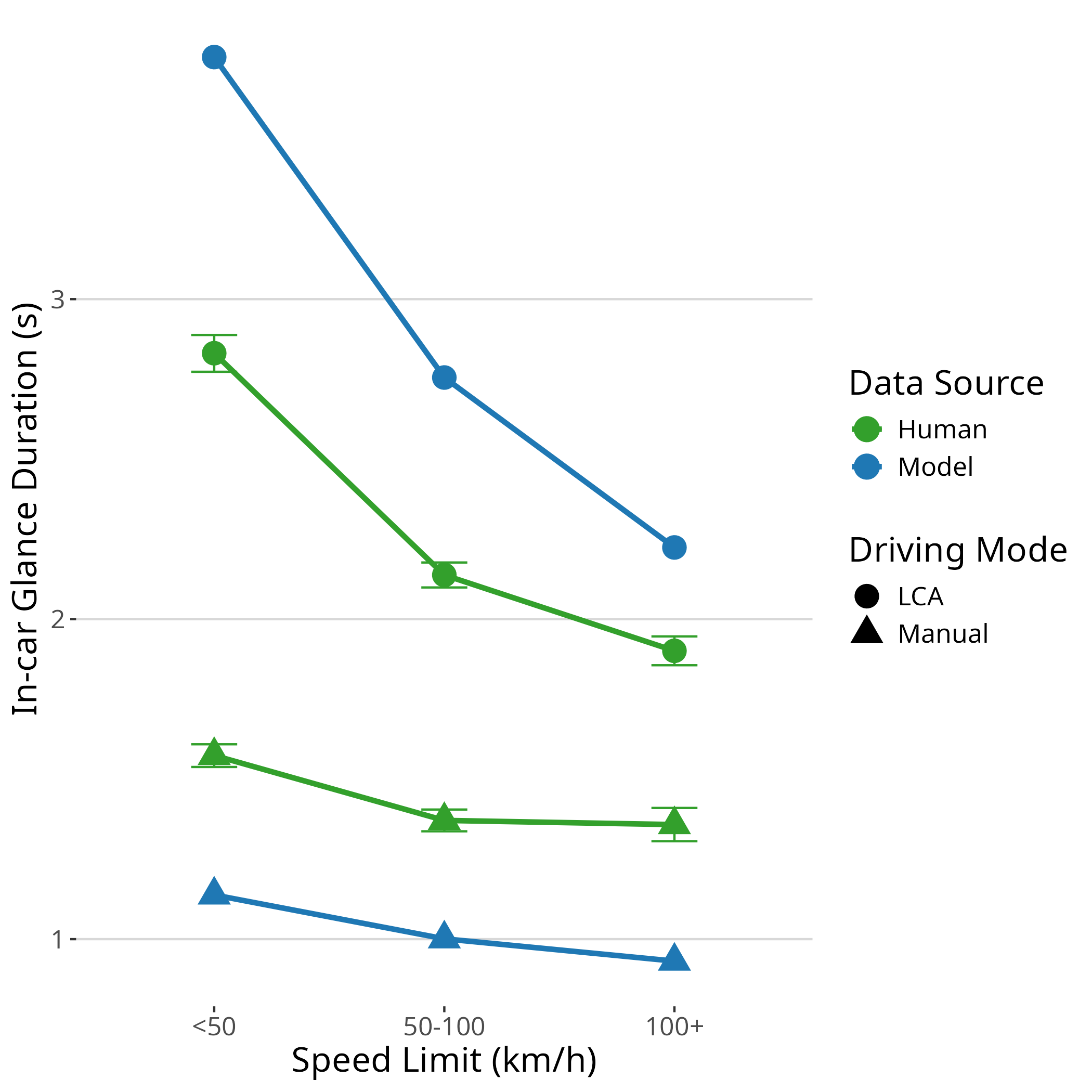}
  \caption{Comparison of our model's prediction of in-car glance durations and those collected from naturalistic driving. Visible is the impact of speed and lane centering assist (LCA) on in-car glance duration.
    The error bars for human data are estimated 95\% confidence intervals.
  }
\label{fig:fit2}
\end{figure}

Our multitasking model replicated the dataset by varying speed limits between 30 and 150 km/h and toggling LCA ``on'' or ``off''.
The number of elements in the in-car search task was also varied between 6 and 12.
Figure~\ref{fig:fit2} shows the model's predictions of the impact of driving speed (divided into three speed bins) and LCA status (on/off) on in-car glance duration.
The model fit was $R^2 = 0.97$, RMSE = 1.7.
The model captures both main effects: in-car glances shorten as speed increases, and glances lengthen when LCA is active.
In absolute terms, the model slightly overestimates glance durations with LCA and underestimates them without LCA.
However, since the human data were collected in a naturalistic environment, including other traffic and, at lower speeds, pedestrians, such discrepancies are expected.

\section{Discussion and Conclusion}

In technology-rich environments, understanding how human cognition adapts to complex interactions and automation is crucial, especially in safety-critical contexts.
Driving, despite its long history, is undergoing a major transformation due to advanced in-car interactions and the spread of (partially) automated driving features.
The safety implications of these technologies depend on our ability to understand human cognition and its adaptability to new systems~\cite{gold2018modeling,dunn2021investigating,lin2018interview,NHTSA2012,Regan.2022,Young.2010}.
In this paper, we develop a computational cognitive model of driver multitasking and argue that such modeling is essential for designing safety-critical systems.

Earlier models of driver interactions have captured certain aspects of driver behavior but often overlook the dynamic adaptability of human cognition when balancing the conflicting goals of safe driving and efficient in-car interaction~\cite{bianchi2020drivers,brumby2007cognitive,gold2018modeling,Horrey2006,jokinen2021multitasking,jokinen2021modelling,kujala2015modeling,lee2017,lee2019modeling,Purucker.2017,salvucci2006modeling,sharp2000mathematical,wierwille1993initial}.
Driving conditions, task design, and varying levels of automation influence how humans adapt their multitasking.
A computational rational approach is particularly suitable for modeling driver multitasking, as it provides a principled framework for predicting adaptive behavior in complex, dynamic environments \cite{oulasvirta2022computational}.
Unlike purely data-driven models, which require extensive empirical datasets and may struggle with generalizability, computational rational models integrate cognitive constraints and task demands to generate behavior aligned with human decision-making processes.
This approach allows for predictions beyond observed data, enabling researchers to explore hypothetical scenarios, assess design trade-offs, and develop more effective in-car interactions.

The computational model developed in this paper formalizes multitasking as adaptation by simulating the effects of road scenarios, in-car task design, and driving automation on multitasking behavior.
Findings such as shorter in-car glance durations on curves compared to straight roads—and how this interacts with speed and automation—demonstrate the model’s ability to generate real-world predictions.
Empirical validation supports the model’s utility, offering both theoretical and practical implications for in-vehicle HCI.
In this paper, our focus was on understanding the cognitive process associated with driver multitasking.
Future research should explore more realistic driving simulators, such as Carla~\cite{Dosovitskiy17} or the coupled simulator~\cite{bazilinskyy2020coupled}.
These simulators offer rich, realistic environments while limiting modeler degrees of freedom.
A key challenge, however, is translating the complex information from such 3D simulators into representations suitable for cognitive models like our multitasking framework.

Understanding the impact of in-car technologies on road safety is crucial.
Specific road conditions, combined with certain in-car tasks, can influence driver risk levels.
Our model confirms these effects by identifying patterns such as longer glances when driving straight, shorter glances when cornering, and longer glances for more complex in-car tasks.
Additionally, the model predicts longer glances when automation, such as lane centering, is active, suggesting a shift in drivers' attention allocation strategies.
While driving automation features may enhance safety, they can also divert attention from potential hazards or reduce drivers' ability to assess whether the vehicle is handling them appropriately~\cite{wintersberger2019overtrust,bianchi2020drivers,dunn2021investigating,lin2018interview}.
Evaluating these behavioral changes is essential to ensure that new technologies do not inadvertently compromise safety.
We release the model code as open source, facilitating further development of our approach.

\bibliographystyle{ACM-Reference-Format}
\bibliography{main}

\end{document}